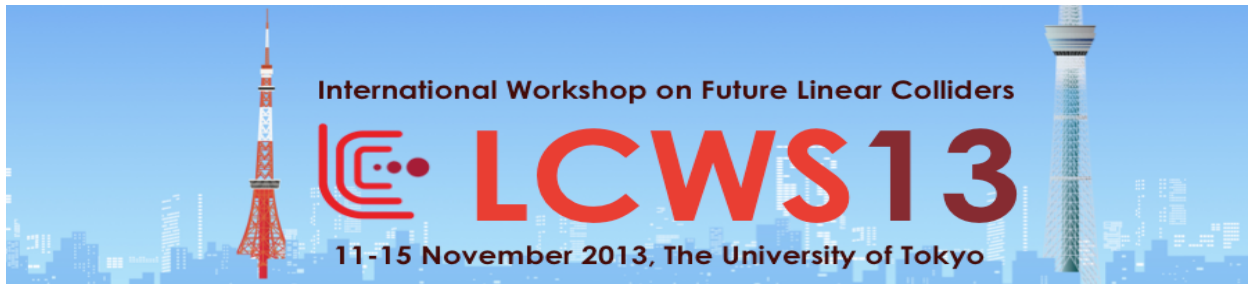

# CALICE Digital Hadron Calorimeter: Calibration and Response to Hadrons

Burak Bilki[1]
on behalf of the CALICE Collaboration

*Argonne National Laboratory, Argonne, IL 60439*
*University of Iowa, Iowa City, IA 52242*

### Abstract

The large CALICE Digital Hadron Calorimeter prototype (DHCAL) was built in 2009 - 2010. The DHCAL uses Resistive Plate Chambers (RPCs) as active media and is read out with 1 x 1 $cm^2$ pads and digital (1 - bit) resolution. With a world record of about 0.5M readout channels, the DHCAL offers the possibility to study hadronic interactions with unprecedented spatial resolution. This talk reports on the results from the analysis of pion events of momenta between 2 to 60 GeV/c collected in the Fermilab test beam with an emphasis on the intricate calibration procedures.

---

[1] Corresponding author: burak-bilki@uiowa.edu

# 1. Introduction

The CALICE Collaboration develops calorimeters that are optimized for the application of Particle Flow Algorithms (PFAs) for future linear colliders [1]. The large CALICE Digital Hadron Calorimeter (DHCAL) prototype was built in 2009-2010. The design of the DHCAL was based on the preliminary work done with a small-scale prototype, which underwent a rigorous test program in the Fermilab test beam and resulted in numerous publications [2-7]. The active media of the DHCAL are Resistive Plate Chambers (RPCs), which are read out by 1 x 1 $cm^2$ pads with a 1-bit resolution (digital readout). A single layer of the DHCAL measures roughly 1 x 1 $m^2$ and consists of 96 x 96 pads. During the Fermilab beam tests, up to 52 layers were installed. The calorimeter consisted of a 38-layer structure (main stack) with 1.75 cm thick steel absorber plates and a 14-layer structure (tail catcher) with eight 2 cm thick steel plates followed by six 10 cm thick steel plates. In addition to the absorber plates, each layer of RPCs was contained in a cassette with a 2 mm thick Copper front plate and a 2 mm thick Steel back plate. The details of the DHCAL are given in [8, 9].

The DHCAL is a calorimeter with the following unique features:
- RPCs for calorimetry (no other hadron calorimeter uses RPCs as active medium),
- Pad readout of RPCs (RPCs are usually readout with strips),
- Digital readout,
- Embedded front-end electronics,
- Large channel count (a world record of ~ 0.5M channels).

Here, we describe the basics of the intricate calibration procedures and their implementation in the analysis of the Fermilab data. Calibration of the W-DHCAL with the CERN test beam data is presented in [10], software compensation techniques for the Fe- and W-DHCAL are introduced in [11] and the recent DHCAL developments are described in [12].

# 2. Calibration Parameters

The DHCAL data contain the hit position information, the time stamp of the individual hits and the time stamp from the trigger and timing unit. Additionally, discriminated signals from a beam Čerenkov counter and a muon tagger (a downstream scintillator 1 x 1 $m^2$ paddle) are integrated into the data stream by the data acquisition system.

The hits in each layer are combined into clusters using a nearest-neighbor algorithm. If two hits share a common edge, they are assigned to the same cluster. The cluster's $x$ and $y$ coordinates are calculated as the average of these coordinates over the constituent hits. Here, the $x$ axis is the horizontal, the $y$ axis is the vertical and the $z$ axis is along the beam direction (a right-handed coordinate system with the origin at the center of the most upstream layer). The event selection requires at least five active layers (layers with at least one hit) in order to eliminate events with spurious triggers.

The calibration of the DHCAL involves several steps. To begin, the performance parameters of the individual RPCs, i.e. the efficiency and the average pad multiplicity, are measured. Here two methods are used: track fits and track segment fits. Dead or hot cells, if any, are identified on a run-by-run basis. In order to avoid a bias in the estimation of the performance parameters, regions within 1 cm of dead/hot cells or RPC edges are excluded from these measurements.

In a second step, the number of hits measured in a given RPC is corrected for differences in its performance parameters. As explained in Section 3, three different approaches have been explored: full calibration, density-weighted calibration, and hybrid calibration.

This talk describes the calibration procedures leading to uniform RPC performance by correcting for the differences between the performance characteristics of the individual RPCs.

## 2.1. Track Fits

The track fits method uses dedicated muon calibration runs to assess the performance parameters of individual RPCs. This method starts with grouping the clusters that are laterally within a distance of 3 cm of each other in different layers. At least one cluster in the first three and one cluster in the last three layers of the main stack is required. If the cluster size exceeds 4 hits in any two consecutive layers, the event is not used for calibration purposes. This selection is to exclude events with interactions within the DHCAL. The tail catcher sections are measured if a valid track fit is performed in the main stack and there are at least three active layers with no interactions in a given tail catcher section.

The group of clusters is then fit to the 3-dimensional parametric line $x=x_0+a_x t$; $y=y_0+a_y t$; $z=t$. $\Delta r/\Delta z$ of the track, where $\Delta r=\sqrt{\Delta x^2+\Delta y^2}$, is required to be less than 0.5 pads/layer and the fit $\chi^2$/ndf is required to be less than 1 (for simplicity, the errors on the cluster positions were taken as 1). For each layer, clusters within 2 cm of the point predicted by the fit are searched for. If a cluster is found, the layer is counted as efficient, and inefficient otherwise. If the layer is efficient, the pad multiplicity is given by the size of the found cluster. If multiple clusters are found in this search, the pad multiplicity is given by the size of the cluster that is closest to the fit point.

## 2.2. Track Segment Fits

The track segment fits method is developed to measure the calibration parameters using the track segments within hadronic showers. With this method, the DHCAL provides another unique feature in calorimetry: For operation in a colliding beam environment, the DHCAL does not need a dedicated calibration system, as track segments can be used to monitor the performance of the RPCs.

The method starts with searching for four clusters that are aligned within 3 cm in four different layers (pick layers). Each of these clusters is required to contain at most four hits, and to be isolated within a radius of 4 cm (no other clusters within 4 cm in the same layer). The track segment is then fit to the parametric line defined in Section 2.1, $\Delta r/\Delta z$ of the track segment is required to be less than 0.5 pads/layer and the fit $\chi^2$/ndf is required to be less than 1, as in the case of the track fits.

This track segment is used to measure the performance parameters of a fifth layer (measurement layer). The measurement layer can either be within the layer span of the pick layers or outside, but only one measurement layer per track segment is allowed. In the measurement layer, clusters within 2 cm to the fit point are searched for. If a cluster is found, the layer is measured as efficient, and inefficient otherwise. If the layer is efficient, the pad multiplicity is measured as the size of the found cluster.

# 3. Calibration Procedures

Using the methods in Section 2, the calibration factors per RPC per data taking run are obtained as $C_i = \varepsilon_i \mu_i / \varepsilon_0 \mu_0$ where $\varepsilon_i$ and $\mu_i$ are the efficiency and the average pad multiplicity of RPC $i$ and $\varepsilon_0$ and $\mu_0$ are the average RPC efficiency and pad multiplicity of the entire stack, 0.96 and 1.56 respectively [9].

Three different calibration procedures are defined and applied to the data:

**Full Calibration:** The hits in RPC $i$ are weighted by $1/C_i$.

**Density-Weighted Calibration:** This approach takes into account that pads collecting charge from several nearby avalanches, for instance in the core of a shower, require a different calibration procedure than pads measuring single tracks. In other words, a pad in the core of a shower will register a hit with minimal dependence on the performance characteristics of this particular RPC and it should be calibrated in a different way than a pad e.g. along a MIP track in the same RPC. In this approach, the calibration factors of a pad may, in general, depend on the local hit density (as a measure of the number of avalanches contributing to the signal charge of that pad), the energy of the incident particle, and the type of incident particle, in addition to the performance parameters of the RPCs.

In order to study this calibration scheme, simulated pion ($\pi^+$) and positron ($e^+$) samples were generated and digitized using the RPC_sim program [9]. Starting with the Geant4 output, the RPC_sim program emulates the response of the RPCs by generating and distributing the avalanche charges, and applying a threshold to the accumulated charge in each pad in order to reconstruct the hits (digitization). To mimic the effect of a specific efficiency and average pad multiplicity of a given RPC, MC samples are digitized with a set of different thresholds that cover reasonably large range of RPC performances. For example, given two different thresholds ($T_1$ and $T_2$), the same Geant4 MC sample is digitized to correspond to two different sets of performance parameters (setting 1: $\varepsilon_1$-$\mu_1$ and setting 2: $\varepsilon_2$-$\mu_2$). The density-weighted calibration method is developed to obtain a correction that is able to correct the setting 1 conditions into the setting 2 conditions.

In a first step, the hits in an event are classified into density bins, where density bin i contains all hits which count i hits in the 3 x 3 array surrounding this hit. The correction factors are determined for each density bin separately. The calculation of the correction factors for the transition from condition 1 to 2 use the fact that the hits in the two digitized samples are correlated.

In a second step, each hit in sample 2 is associated with the density bin of its correlated hit in sample 1. If there is no one-to-one correlation for a given hit in sample 2, the hit is associated with the geometrically closest hit in sample 1.

In a third step, the calibration factor for density bin i is determined as the average ratio over the whole sample of the number of hits in sample 2 correlated to hits in density bin i of sample 1 to the number of hits in density bin i of sample 1. These calibration factors can take values both smaller and larger than unity e.g. if sample 2 is digitized with a larger average pad multiplicity, then additional hits will be correlated to hits in a given density bin of sample 1, resulting in calibration factors larger than one.

Figure 1 shows an example of utilization of the correction factors for a 10 GeV pion sample starting from $T_1$=400 (arbitrary charge units roughly corresponding to 1 fC) to reproduce the response corresponding to $T_2$=800. The correction factors determined with the procedure described above are shown in Fig. 1c. Figure 1a shows the simulated response at $T_1$. Figure 1b is the response after the correction factors are applied and Fig. 1d shows the original response at $T_2$. As a result of this calibration procedure, the average response and energy resolution at $T_2$ are precisely reproduced. Similar results are obtained when trying to recover lower threshold digitized samples from higher threshold ones. The average correction factors are found to exhibit only a weak energy dependence. Hence, a single calibration procedure for all energies is possible.

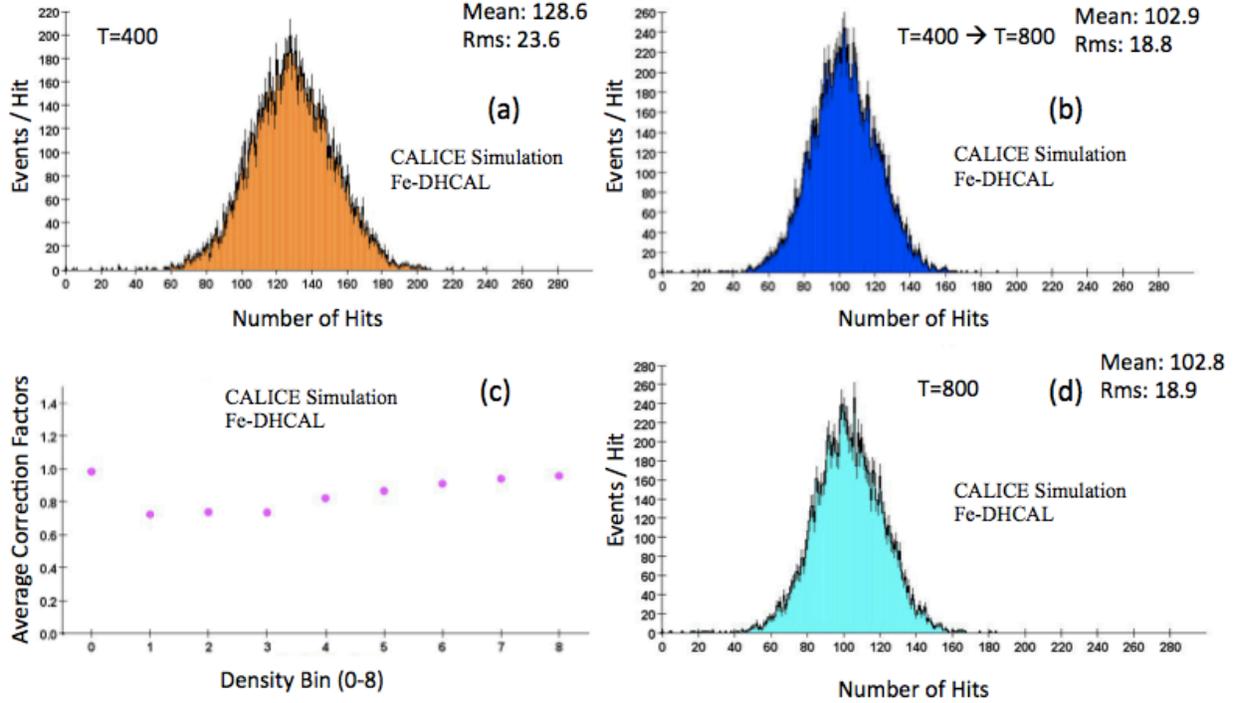

**Figure 1.** Demonstration of the density-weighted calibration procedure using 10 GeV pion MC sample digitized with two different thresholds $T_1$=400 and $T_2$=800. (a) Response at $T_1$=400, (b) corrected response at $T_1$=400 to obtain the response at $T_2$=800, (c) average correction factors as a function of the density bins (0-8) and (d) the original response at $T_2$=800.

The average correction factors cannot be determined from data, since data cannot provide simultaneously two different operating conditions for a given RPC. The density-weighted calibration only depends on the set of four calibration parameters $\varepsilon_i$, $\mu_i$, $\varepsilon_0$ and $\mu_0$, where $\varepsilon_i$ and $\mu_i$ are the performance parameters of RPC $i$ and $\varepsilon_0$ and $\mu_0$ are the average performance parameters of the stack.

For each density bin separately, the correction factors $C$ were plotted as a function of $R = \varepsilon_i \mu_i / \varepsilon_0 \mu_0$. However, it was found that the correction factors can not be parameterized by a smooth function of this definition of $R$. Better results were obtained empirically with $R_\pi^+ = \varepsilon_i^{0.3} \mu_i^{1.5} / \varepsilon_0^{0.3} \mu_0^{1.5}$ and $R_e^+ = \varepsilon_i^{0.3} \mu_i^{2.0} / \varepsilon_0^{0.3} \mu_0^{2.0}$, where the subscript denotes the particle type. Figure 2 shows the correction factors as a function of $R$. Each point in these plots is the average correction factor obtained as described above for a different set of performance parameters. These factors were fit to a power law $C = p_0 R_p^{p_1}$ with parameters $p_0$ and $p_1$. In the case of density bin 0, a constant was added to the function, $C$ $= p_0 R_p^{p1} + p_2$, in order to obtain a satisfactory fit. The empirical exponents in $R$s were obtained by minimizing the $\chi^2$'s of the fits. Therefore, Fig. 2 defines the entire density-weighted calibration procedure as a function of the beam type, performance parameters and the densities.

The utilization of the correction factors is as follows: For a given hit in an RPC with performance parameters of $\varepsilon_i$ and $\mu_i$, $R$ is calculated depending on the showering particle type. Then, the correction factor $C$ is obtained using the fit function of the density bin that this hit belongs to. Finally, the hit is weighted by $C$.

**Hybrid Calibration:** For the hits with 0 or 1 neighbor, the density effect is minimal. The hybrid calibration utilizes full calibration for density bins 0 and 1, and density-weighted calibration for the higher density bins.

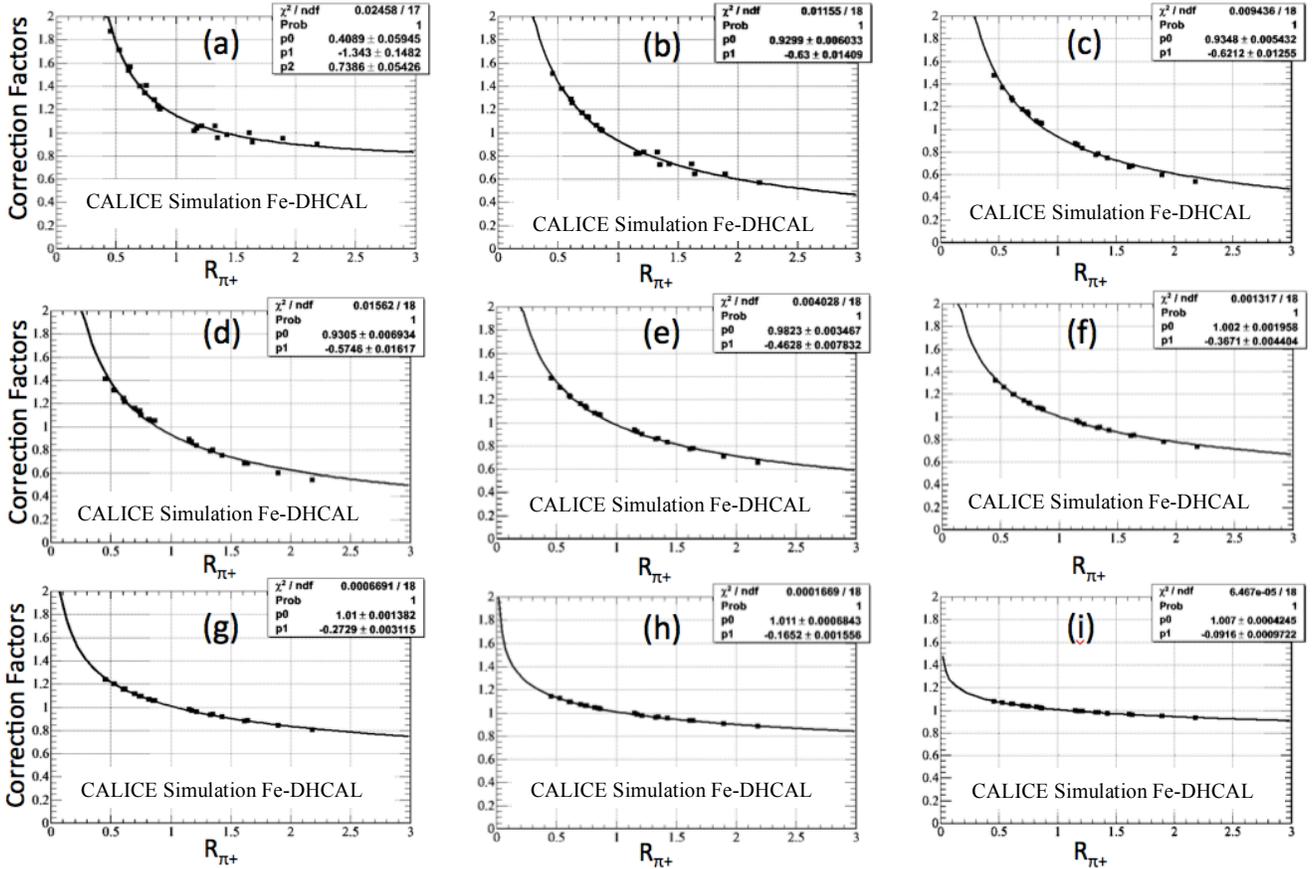

**Figure 2.** $\pi^+$ correction factors for density bins of 0–8 (a–i). The points are fit to power functions. $R_{\pi^+} = \varepsilon_i^{0.3} \mu_i^{1.5} / \varepsilon_0^{0.3} \mu_0^{1.5}$.

Figure 3 shows the application of the three calibration schemes to the 4 GeV $\pi^+$ (left) and 8 GeV $e^+$ (right) data. As expected, the uncalibrated data (black) show the largest amount of fluctuation in response between different runs. The full calibration (red), density-weighted calibration (green) and the hybrid calibration (blue) schemes all result in improved uniformity of the responses with significantly smaller fluctuations. All calibration schemes are successful in compensating for the slight differences in the RPC performance characteristics.

Figure 4 shows the $\chi^2$/ndf for the constant line fits to the $\pi^+$ data collected in the Fermilab test beam. All calibration schemes improve the uniformity of the response across different runs and run periods. The three calibration schemes seem to perform at similar levels with no clear winner.

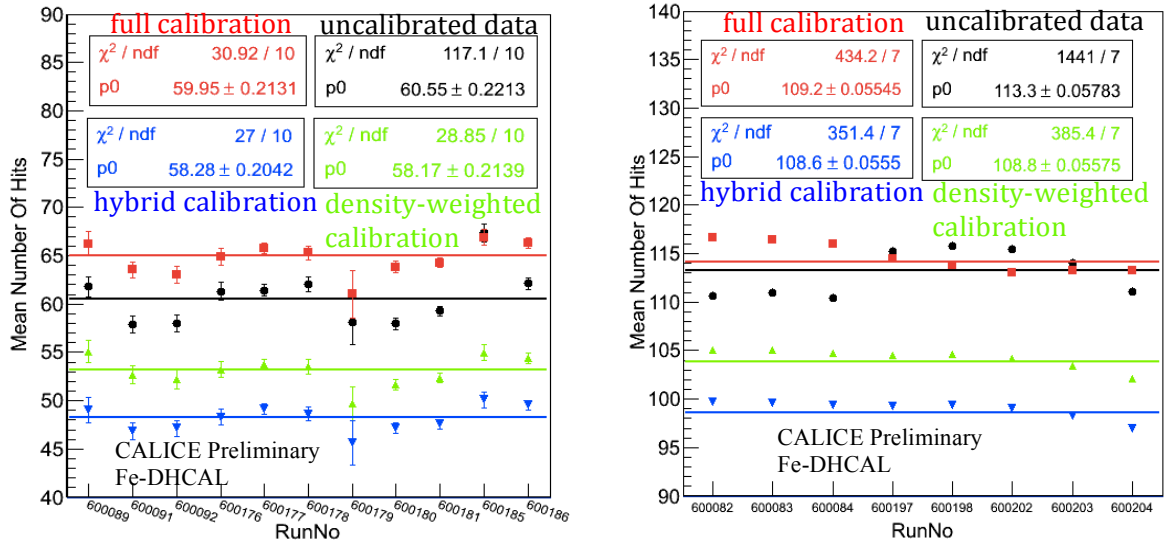

**Figure 3.** The results of the three calibration schemes applied to the 4 GeV $\pi^+$ (left) and 8 GeV $e^+$ (right) data. The uncalibrated data (black, 0), full calibration (red, 5), density-weighted calibration (green, -5) and the hybrid calibration (blue, -10) responses are all fit to a constant. The numbers in the parenthesis following the colors are the y-offsets applied to the data points in order to increase their visibilities. The error bars show the statistical errors on the means.

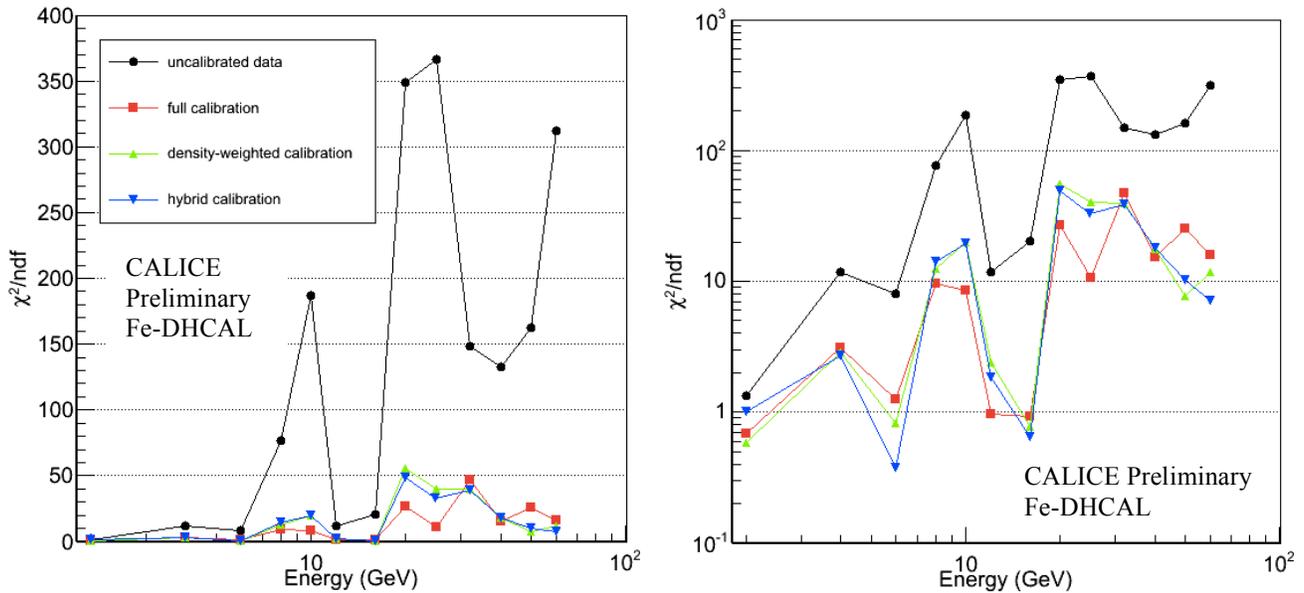

**Figure 4.** Normalized $\chi^2$ of the fits of a constant to all the $\pi^+$ runs at a given energy (left), also in log-y scale (right).

# 4. Pion Results

The event selection requires not more than 1 cluster in Layer 1 with at most four hits. This selection assures that upstream interactions are not included. The requirement of at least five active layers rejects events with spurious triggers.

In order to separate the pions, positrons and muons in the beam, the Čerenkov counter in the beamline is utilized. In addition, a topological particle identification (PID) method is developed and implemented for 2, 4, 25 and 32 GeV data where the Čerenkov counter was not efficient. For details of the topological PID see [13].

The RPC performance parameters are calculated using the track segment fits method.

Figure 5 shows the mean response (a, c, e) and the energy resolution (b, d, f) for the uncorrected pion data (black in all plots), full calibration (red in a, b), density-weighted calibration (green in c, d) and hybrid calibration (blue in e, f). The mean response is fit to the power function N=aE$^m$ up to and including 60 GeV. The resolutions are fit to the generic $\frac{\sigma(N)}{N} = \frac{\alpha}{\sqrt{E}} \oplus C$ where $\alpha$ is the stochastic term and C is the constant term. The resolution fits are up to and including the 25 GeV point and they are extrapolated to 60 GeV. No additional corrections/selections are applied to the data (e.g. containment cuts, correction for response non-linearity). Therefore, the purpose of Fig. 5 is to demonstrate the effect of the calibration schemes on the results.

All calibration schemes tend to normalize the mean response to the predefined DHCAL operating conditions. At lower energies, the methods agree with each other. However, at higher energies where the shower densities are large, the effect of employing the density weighting in the calibration procedure is clearly visible.

Figure 6 shows the percent deviation between the mean response fit values (Fig. 5 a, c, e) and the data points for both the uncalibrated data and the calibrated data. All calibrated responses have a deviation less than ±4 % at all energies except at 2 GeV. The systematic error associated with the calibration procedures is estimated to be less than 0.1 %. This estimate is based on the performance studies of the density-weighted calibration using MC samples (see e.g. Fig. 1). No other systematic effects are discussed in this paper (e.g. the uncertainties in the beam energy, the variations in the beam position and angle, the particular location of dead areas, etc.).

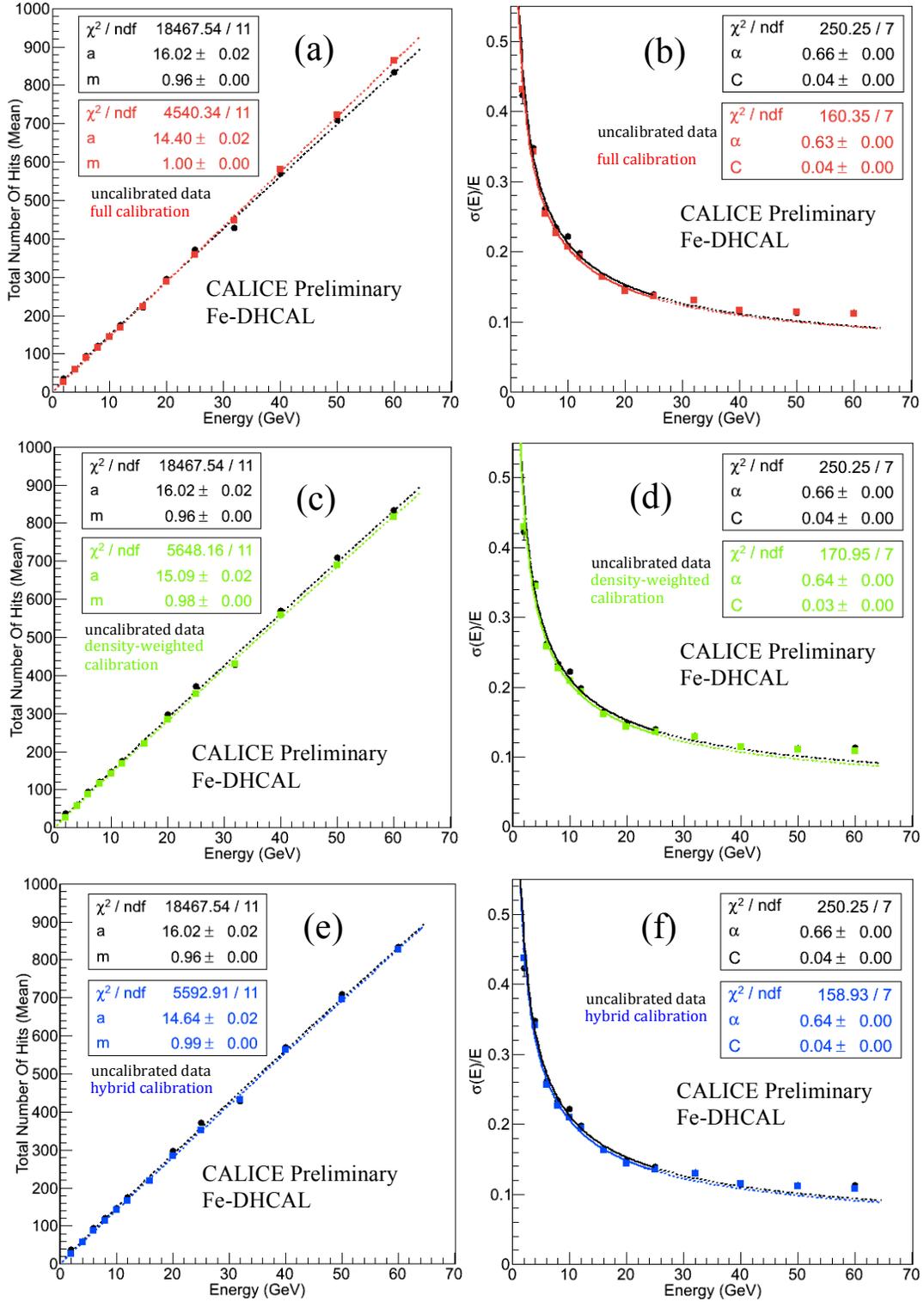

**Figure 5.** Mean response (a, c, e) and resolution (b, d, f) for the uncalibrated pion data (black) and the three calibration schemes (full calibration – red; density-weighted calibration – green; hybrid calibration – blue). For all calibration schemes, the fit quality is improved both for mean response (a, c, e) and resolution (b, d, f) compared to the fits to the uncalibrated data. The resolution fits (b, d, f) are up to 25 GeV (solid) and are extrapolated to 60 GeV (dashed).

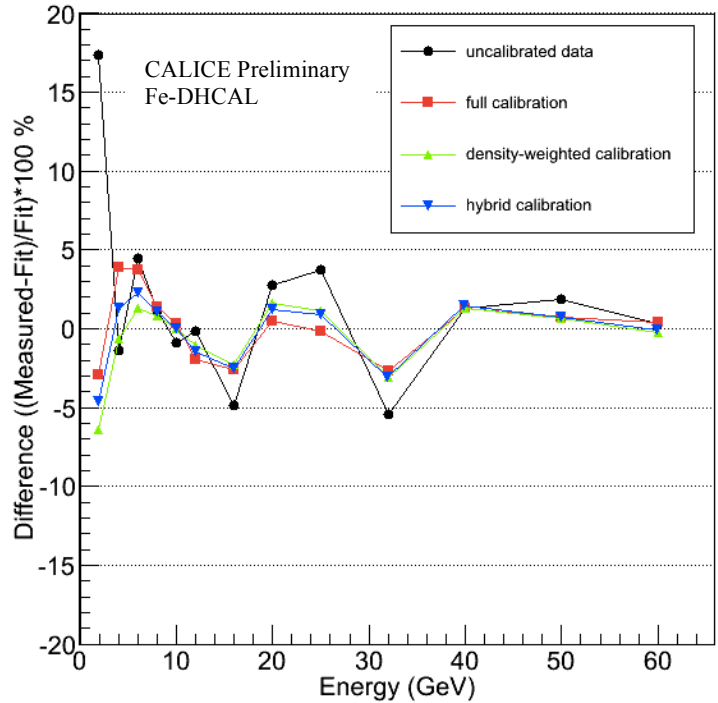

**Figure 6.** Percent difference between the mean response fit values (Fig. 5 a, c, e) and the data points for the uncalibrated data (black) and the calibrated data (full calibration – red; density-weighted calibration – green; hybrid calibration – blue).

## 5. Conclusions

The DHCAL recorded around 14 million secondary beam events over five test beam campaigns at Fermilab. The beam is a momentum-selected mixture of muons, pions and positrons. Data were collected at 2, 4, 6, 8, 10, 12, 16, 20, 25, 32, 40, 50, 60 GeV. The high granularity and the digital readout of the DHCAL enable the utilization of numerous topological event parameters for all purposes ranging from calibration to correcting the hadronic/electromagnetic response (software compensation) and improvements to the energy resolution measurements.

The calibration of the DHCAL is based on two performance parameters of the Resistive Plate Chambers: efficiency and average pad multiplicity. To first order, a simple multiplication of these parameters normalized to a reference value can serve as a calibration factor. However, the density of showering particles per pad impacts the calibration procedure in a complicated manner. As a result, the density-weighted calibration schemes provide better handles in understanding/manipulating response differences due to changes in individual RPC performances and operation conditions of the DHCAL.

All three calibration schemes (full calibration, density-weighted calibration and hybrid calibration) result in a more uniform response for all runs at each energy point when compared to the uncalibrated results.